\documentclass{aa}

\usepackage{graphicx}
\usepackage{txfonts}

\def\Msun{M$_{\sun}$}
\def\Mjup{M$_{\rm Jup}$}

\def\bp{$\beta$~Pic}
\def\au{{\sc au}}
\def\Ksabs{$M_{Ks}$}

\def\Lpabs{M$_{L'}$\ }
\def\L'{$m_{L'}$\ }
\def\cc{cc}
\renewcommand{\d}{\textrm{d}}
\begin{document}

\title{Constraining the orbit of the possible companion to $\beta$~Pictoris \thanks{Based on observations collected at the
European Southern Observatory, Chile, ESO; runs 282.C5037(A), 282.C5037(B) and 282.C5037(D).}}
\subtitle{New deep imaging observations}

\author{
  A.-M.~Lagrange \inst{1}
  \and
  M.~Kasper \inst{2}
  \and
  A. Boccaletti \inst{3}
  \and
  G.~Chauvin \inst{1}
  \and 
  D.~Gratadour \inst{3}
  \and
  T.~Fusco \inst{4}
  \and
  D.~Ehrenreich \inst{1}
  \and 
  D. Apai \inst{5}
  \and 
  D.~Mouillet \inst{1}
  \and
  D.~Rouan \inst{3}
}

\offprints{A.-M. Lagrange}

\institute{
  Laboratoire d'Astrophysique de l'Observatoire de Grenoble, Universit\'e Joseph Fourier, CNRS (UMR~5571), \\
  BP 53, 38041 Grenoble, France, \email{anne-marie.lagrange@obs.ujf-grenoble.fr}
  \and
  European Southern Observatory, Karl Schwarzschild Stra\ss e, 2, 85748 Garching bei M\"unchen, Germany
  \and
  Laboratoire d'\'Etudes Spatiales et d'Instrumentation en Astrophysique, Observatoire de Paris, CNRS (UMR~8109), Universit\'e Pierre et Marie Curie, Universit\'e Paris-Diderot, 5, place Jules Janssen, 92195 Meudon, France
  \and
  Office National d'\'Etudes et de Recherches A\'erospatiales, 29, avenue de la Division Leclerc, 92322 Ch\^atillon, France
  \and 
  Space Telescope Science Institute, 3700 San Martin Drive, Baltimore, MD 21218, USA
}

\date{Accepted in Astronomy \& Astrophysics.}

\abstract
{We recently reported on the detection of a possible planetary-mass companion to \object{$\beta$~Pictoris} at a projected separation of 8~\au\ from the star, using data taken in November 2003 with NaCo, the adaptive-optics system installed on the Very Large Telescope UT4. Eventhough no second epoch detection was available, there are strong arguments to favor a gravitationally bound companion rather than a background object. If confirmed and located at a physical separation of 8~\au, this young, hot ($\sim 1500$~K), massive Jovian companion ($\sim 8$~\Mjup) would be the closest planet to its star ever imaged, could be formed via core-accretion, and could explain the main morphological and dynamical properties of the dust disk.} 
{Our goal was to return to \bp\ five years later to obtain a second-epoch observation of the companion or, in case of a non-detection, constrain its orbit.}
{Deep adaptive-optics $L'$-band direct images of \bp\ and $K_s$-band Four-Quadrant-Phase-Mask (4QPM) coronagraph images were recorded with NaCo in January and February 2009. We also use 4QPM data taken in November 2004.}
{No point-like signal with the brightness of the companion candidate (apparent magnitudes $L'=11.2$ or $K_s \simeq 12.5$) is detected at projected distances down to $\simeq 6.5$~\au\ from the star in the 2009 data.} 
{As expected, the non-detection does not allow to rule out a background object; however, we show that it is consistent with the orbital motion of a bound companion that got closer to the star since first observed in 2003 and that is just emerging from behind the star at the present epoch. We place strong constraints on the possible orbits of the companion and discuss future observing prospects.}


\maketitle

\section{Introduction}
The  $\beta$~Pic disk of dust and gas has been regarded as the prototype of 
young ($12^{+8}_{-4}$~Myr; \cite{zuckerman01}) planetary systems since the 1980's and has revealed over the years an impressive amount of indirect
signs pointing toward the presence of at least one giant planet.

Using $L'$-band high-angular resolution imaging data obtained with the NAOS-CONICA adaptive optic system (NaCo) on the Very Large Telescope UT4, we discovered a point-like source with an apparent magnitude $L' = 11.2\pm0.3$~mag at $0\farcs411 \pm 0\farcs008$ ($\sim 8$~\au) north-east of \bp, well aligned with the dust disk. Eventhough no second epoch data were available with a sensitivity enough to detect the companion candidate (\cc), we were confident that this signal was not due to a background object as the associated probability was shown to be very low. We therefore attributed the source to a {\it probable} bound object. Given the star proper motion, a background object would lie now angularly too close to the star to be detected again and a few more years (see below) are necessary to rule out this possibility from an observational point of view.

Very interestingly, if this \cc\ is indeed at a physical separation of 8~\au, it would explain most of the \bp\ system's morphological and dynamical peculiarities: the disk inner warp, its brightness asymetries, as well as the observed falling evaporating bodies (FEBs; \cite{lagrange09}). Recently, Lecavelier des Etangs \& Vidal-Madjar (2009) investigated whether the observed \cc\ could also be responsible for the photometric variability observed in November 1981 (\cite{leca95}) and analysed in details in a subsequent paper (\cite{leca97}). In the latter paper, the complex photometric curve observed was explained either by a planet located in a dust-free region of the disk, or a cloud of dust passing in front of the star. In the planet scenario, an object with a radius larger than 2 Jupiter radii, located at less than 8~\au\ and surrounded by a void of material within its Hill radius could explain both the eclipse signal and the observed higher brightness a few days around the eclipse. From their dynamical analysis, Lecavelier des Etangs \& Vidal-Madjar (2009) conclude that the \cc\ observed in 2003 could be responsible for the 1981 eclipse provided the semi-major axis of its orbit (circular case or assuming a low eccentricity) is in the range 7.6--8.7~\au, corresponding to periods in the range 15.9--19.5~years. Another longer-period orbit, dynamically compatible with the existence of an eclipse and the November 2003 data was excluded on the basis that the \cc\ would have been detected earlier. We note that the corresponding orbital radius, 17~\au\ is also much larger than the one predicted by the modeling of the photometric curve. As noted by these authors and Lagrange et al.\ (2009), a \cc\ on an orbit with a radius of $\sim 8$~\au\ would not be detectable in fall 2008 nor in early 2009. Lecavelier des Etangs \& Vidal-Madjar (2009) furthermore predict that if the \cc\ is responsible for the  1981 eclipse, then it should reach its maximum elongation between 2011 and 2015. Finally, we note that the data available to Lecavelier des Etangs \& Vidal-Madjar (2009) did not allow them to disentangle two possible cases: a first case where the \cc\ would have been located before quadrature in 2003, and a second case where the \cc\ would have been located after quadrature in 2003.

Obviously, new deep imaging observations are needed to further constrain the possible orbits of the \cc. A major question is the true (unprojected) separation of the \cc\ from the star. Four Quadrant Phase Mask (4QPM) coronagraph observations performed in 2004 did not reveal any point-like source with an absolute magnitude\footnote{In the following, absolute magnitudes in a photometric band $i$ are noted $M_i$ and calculated assuming a distance of 19.3~pc. The notation $i$ will simply refer to the apparent magnitude in this band.} in the $K_s$ band of $M_{K_s} \simeq 11.5$ at a projected separation $>8$~\au\ (\cite{bocca09}). Using the COND and DUSTY models (\cite{baraffe03}; \cite{chabrier00}), we obtained a $K_s - L'$ color of 1.2 (COND) to 1.4 (DUSTY); hence a \cc\ with $L' = 11.2\pm0.3$ (i.e., an absolute magnitude $M_{L'} = 9.8\pm0.3$) would have $M_{K_s} = 11.0\pm0.3$ to $11.2\pm0.3$, respectively. If located beyond 8~\au, the \cc\ seen in 2003 would then have been detected on these 4QPM data. This allows to conclude that its  projected separation had decreased between 2003 and 2004. 

In order to confirm the companionship and/or to further constrain the \cc\ orbit, we obtained discretionary time to perform in January and February 2009 new high-contrast and high-spatial-resolution observations of the \bp\ system with NaCo (\cite{rousset03}; \cite{lenzen03}) at $L'$ band, allowing thus a direct comparison between 2003 and 2009 data, as well as with the 2004 $Ks$-band 4QPM data.\footnote{Note that we additionally acquired images of \bp\ using the new Sparse Aperture Masking mode offered on NaCo in November 2008. However, the experimental observing template we tested as well as the SAM capabilites do not provide constraints on the \cc. They are consequently not considered here.} The observations were designed to detect the faint companion as close as possible to the star. A positive detection would both confirm the companionship and provide crucial constraints on the companion orbital parameters. A non-detection would provide valuable constraints on the orbit of the \cc. We present these observations in Section~2, and we use these new results to constrain the possible location of the \cc\ in Section~3.

\section{Observations and data reduction procedures}
\subsection{$L'$- band observations (February 2009)}
\label{sec:obs} 
\subsubsection{Observing strategies}
$L'$-band images of \bp\ ($V=3.8$, $L'=3.5$) were obtained in February 2009 with NaCo. Two different observing strategies were adopted. The first one (run A) followed as closely as possible the observing procedure adopted in November 2003, which consisted in recording \bp\ non-saturated and saturated images, followed by similar data on a comparison star, HR~2435 (also used in November 2003), whose saturated images are used to remove the stellar halo on the \bp\ saturated images. 

For the second strategy (runs B1 and B2), \bp\ and HR~2435 images were recorded at two different de-rotator positions, to be able to use either \bp\ or HR~2435 to remove the star halo (see examples of the use of the star itself observed at different rotator positions to remove the PSF halo in \cite{kasper07}). The offset between the two rotator positions was chosen to be 30\degr\ (run B1) or 180\degr\ (run B2). The latter was chosen such that  the VLT aperture spiders remain in the same orientations.

For all runs, the time elapsed between \bp\ and HR~2435 observations was precisely calculated so that the images for both stars were recorded at similar parallactic angles, within 0.5\degr, so as to remove as accurately as possible the PSF wings.

\subsubsection{Instrumental set-up}
The visible wavefront sensor was used with the $14 \times 14$ lenslet array, together with the visible dichroic. We used a set-up similar to the one used for the November 2003 observations: CONICA L27 camera, which provides a pixel scale of $\sim 27$~mas; note, however, that the CONICA detector was changed between both observing runs. Saturated images of \bp\ were recorded, with detector integration times (DITs) of $0.2$~s and number of detector integrations (NDIT) of $150$. For the saturated images, we used a four-position dithering pattern every two $\rm DIT \times NDIT $ exposures. This allows accurate sky and instrumental background removal. 

Non-saturated images were also recorded to get images of the stellar point spread function (PSF) as well as a photometric calibration. In such case, we added the Long Neutral Density filter (transmission $\sim 0.018$) in the CONICA optical path, and recorded images with DITs of $0.2$s. Finally, twilight flat fields were recorded as well. 

The log of the observations is reported in Table~\ref{stats}, as well as the observing conditions. Noticeably the observing conditions were not as good as during the November 2003 run, with coherent energies (estimated in the $K$ band) of 40--50\% for run A, and 25--30\% for runs B1 and B2, instead of 50--70\% in 2003, and coherent times $\tau_{0}$ between 2.8 and 6.2~ms, instead of 20~ms in the best data sets in November 2003. The conditions during run B1 and even more during run B2 were rather poor, resulting in an unstable and mediocre adaptive optics (AO) correction.

\begin{table*}
\caption{Observing log of the \bp\ saturated images, and corresponding atmospheric conditions.}
\label{stats}
\centering
\renewcommand{\footnoterule}{}  

\begin{tabular}{l l l l l l l l l l l ll}
\hline \hline
\multicolumn{10}{c}{}     \\
Set & Star & Date        & Filter & DIT & NDIT & $N_\mathrm{exp} $ & $\pi$ $^1$ & $\langle EC \rangle$ $^2$ & $\langle\tau_0\rangle$ $^3$  \\
    &      &             &        & (s) &      & (s)              & (o)        & (\%)                      & (ms)                         \\
\hline
A & \bp\    & 2009-02-11 & $L'$   & 0.2 & 150 & 48 & $-24.2 $/$-9.1$  & 43.4 & 5.0 \\
  & HR~2435 & 2009-02-11 & $L'$   & 0.2/0.22 & 150 & 48 & $-24.7 $/$-10.4$   & 49.0 & 4.9 \\

\hline
B1& \bp\    & 2009-02-16 & $L'$   & 0.2 & 150 & 44 & $-22.8 $/$-8.4$   & 35.7 & 3.2 \\
  & HR~2435 & 2009-02-16 & $L'$   & 0.2 & 150 & 54 & $-22.9 $/$-6.7$  & 26.9 & 2.8 \\

\hline
B2 & \bp\   & 2009-02-17 & $L'$   & 0.2 & 150 & 42 & $-16.8 $/$-2.4$  & 26.8 & 4.4 \\
  & HR~2435 & 2009-02-17 & $L'$   & 0.2 & 150 & 42 & $-16.9 $/$-2.6$  & 28.0 & 6.2 \\
\hline
D & \bp\    & 2009-01-20 & $K_s$  & 0.3  & 200 & 8 & $-30.9 $/$-21.5$   & 51.2 & 9.6 \\
  & HR~2435 & 2009-01-20 & $K_s$  & 0.3  & 200 & 10 & $-31.3 $/$-18.2$   & 44.2 & 8.7 \\

\hline
\end{tabular}
\begin{list}{}{}
\item[$^1$] Range of parallactic angles at the start/end of the observation. 
\item[$^2$] The average coherent energy (K-band) as estimated on-line by AO during saturated exposures. 
\item[$^3$] The average coherence time as estimated on-line by AO during saturated exposures. 
\end{list}
\end{table*}

\subsubsection{Data processing}
The data were reduced using different methods. The first two methods are described in Lagrange et al.\ (2009). A third methods, which takes profit of the different rotator positions in order to remove the \bp\ PSF halo consists in the following steps. The initial steps of the image processing are standard (flat fielding, bad-pixel cleaning, sky subtraction). As in the previous methods, the sky for each exposure was determined using the exposure of the next adjacent dithering position. Hence, the sky measurement for each exposure is never older than about half a minute. Then, the images were superimposed with sub-pixel accuracy by a cross-correlation technique and added up, flux normalized to unit peak intensity using the flux in the unsaturated Airy rings and high-pass filtered. The high-pass filtering was done by subtracting from the image its median filtered ($\approx 3$ full-width-at-half-maximum box-width) version. This procedure efficiently eliminates the large-scale structures while leaving point sources (such as substellar companions) relatively unaffected. The high-pass filtering reduces the PSF peak intensity by only about 15\%. Finally, images taken at different roll angles of the instrument were superimposed and subtracted from each other in order to calibrate residual speckle noise.

\subsection{$K_s$-band observations (January 2009)}
\subsubsection{Observing strategy and instrumental set-up}
As the brightness ratio of the \cc\ in $K_s$ is expected to be larger than in $L'$, a dedicated observing procedure was considered. Coronagraphic images of \bp\ were recorded in the $K_s$ band with the 4QPM installed in 2007 in the NaCo focal wheel (\cite{boccaletti07}) in combination with the Full Undersized pupil stop (10\% undersizing of the UT pupil). Additionally, a reference star was observed and used to calibrate the residual speckle halo (see below). Two series of exposures with $\rm DIT=0.3$~s and $\rm NDIT=200$ was obtained at 4 different rotator positions: 0, 15, 30 and 45\degr\ clockwise, totalizing 480~s. Photometric references obtained on out-of-mask unsaturated images ($\rm DIT=0.8$~s, $NDIT=20$) are corrected for the Neutral Density transmission ($1/90$) and the difference of pupil stop (\cite{bocca08}). The same observing procedure was repeated on the reference star HR~2435. However, the parallactic angles differs by 0.5, 1.4, 3.5 and 3.3\degr\ at each rotator angle. Differences of more than 0.5\degr\ are not well adapted to our needs.

For these $K_s$-band observations, the visible wavefront sensor again was used with the $14 \times 14$ lenslet array, together with the visible dichroic. We used the CONICA S13 camera, which provides a pixel scale of 13.25~mas. The DCS detector mode and readout modes were respectively set to `HighDynamic' and `Double'.

\subsubsection{Data processing}
After standard flat-field correction and bad pixels removal, the 400 images ($\rm DIT=0.3$~s) obtained at each rotator angle were co-added resulting in 8 images of 120~s for the target star and similarly for the reference star. The stability of the stellar image behind the 4QPM was sufficient and did not cause significant variations of the coronagraphic attenuation, so image selection was not necessary. In addition to twilight flat field calibration, we acquired background observations with the 4QPM in the beam that we used to evaluate the 4QPM flat field alone. Correction of this flat field was made on the 120~s exposures as the subtraction of the IR background evaluated on the histogram of pixels. Images were recentered at the sub-pixel scale using the maximum of the cross-correlation map (we checked that it provides satisfying results on coronagraphic images). Then, several methods have been considered to combine this set of images.

\begin{itemize}
\item{Method 1: Images are de-rotated and co-added, resulting in 2 images one for the target star and one for the reference star. The speckle background is averaged by a factor equal to the number of rotator angles ($N_{rot}$). An optimal subtraction is performed between the star and the reference.}\smallskip
\item{Method 2: Image at rotator position $i+1$ is subtracted to image at position $i$. The resulting subtraction is de-rotated and co-added to the previous one. The same procedure is repeated on the reference star images. Finally, the reference image is subtracted from the target image. The speckle background is here averaged by a factor $N_{rot}-1$.}\smallskip
\item{Method 3: The target star image is subtracted with the reference star image for each rotator angle. As in method 2, this image at position $i+1$ is subtracted to image at position $i$. The final image is obtained after de-rotating and co-adding each of these subtractions. Again, the speckle are averaged by a factor $N_{rot}-1$.}\smallskip
\item{Method 4: The last method is closer to that of Marois et al.\ (2006). A median map is estimated on each pixels for the target set of images as for the reference star. This median map is subtracted to each rotator position. Then, the images are derotated and co-added. As for method~1, the background is averaged by a factor of $N_{rot}$.}
\end {itemize}

Every time a subtraction is made between 2 images, we searched for the intensity ratio for an optimal subtraction by scanning a large range of possible ratios. This intensity ratio is then estimated in different ways (in the whole image, in a disk or in a ring) and the average is taken as the result. In principle, the methods~2 and~3 are more efficient than traditional averaging of derotated images as it is shown in the NaCo user manual. However, in the present case, $N_{rot}=4$ is too small and so the gain of these methods is negligible. The results presented in this paper are obtained with the method~4 although the performance of all the methods are not much different at an angular distance of 0\farcs4.

\section{Results}
\subsection{Detection limits}

\begin{figure}
  \centering  \includegraphics[width=0.95\hsize]{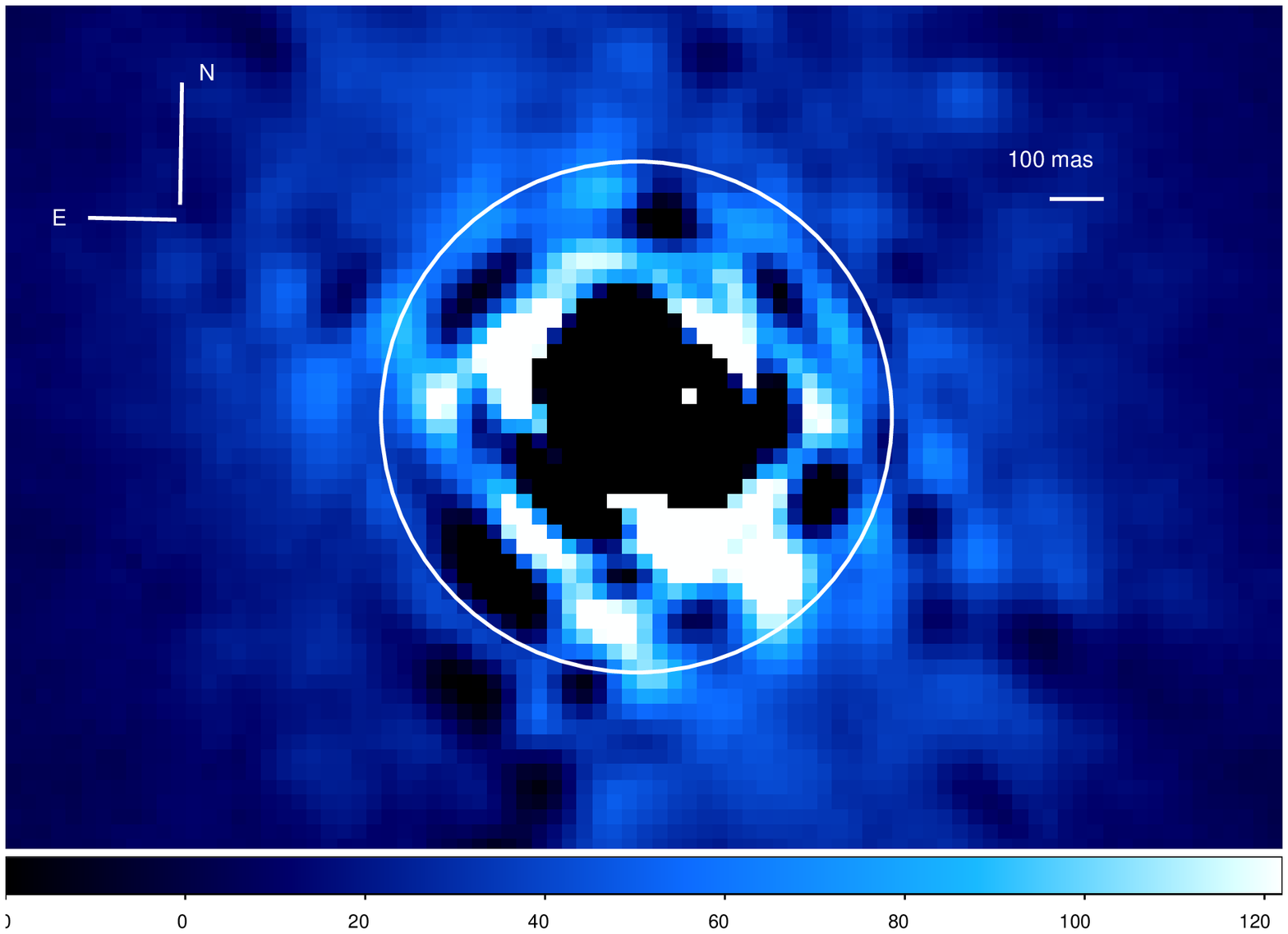}\\ 
  \includegraphics[width=0.95\hsize]{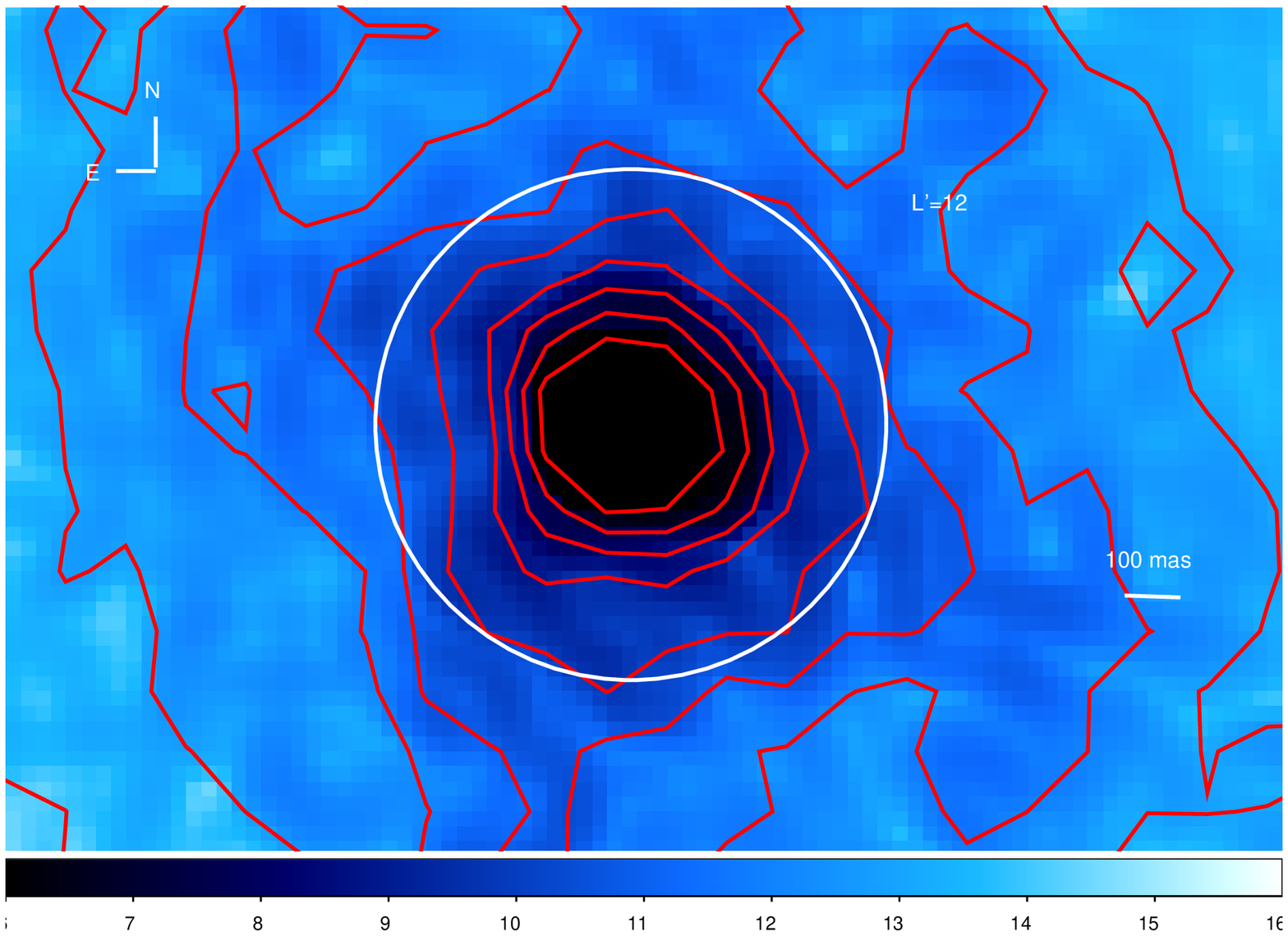} 
  \caption{Top: Residual $L'$ image after subtracting \bp\ by the comparison star HR~2435 (run A data). Bottom: Map of the corresponding $6\sigma$ detection limit. The contour levels represent iso-absolute     magnitudes and are separated by 1~mag. The $M_{L'} = 12$ contour level is indicated for reference. The \cc, with an absolute magnitude $M_{L'} \simeq 9.8$ is not detected in these data. The radius of the white circle corresponds to 8~\au, i.e., the projected separation of the \cc\ in November 2003.}
  \label{images_comp}
\end{figure}

We show in Fig.~\ref{images_comp} the subtracted image corresponding to run A data, as well as the corresponding $6\sigma$ detection limit as obtained when using HR 2435 to remove the PSF halo. The detection limits were computed by measuring for each pixel the noise within a $5 \times 5$-pixel box centered on the given pixel, and determining the corresponding $6\sigma$ limits, using the non-saturated images to get a photometric scaling factor, taking into account the instrumental sets up used to record the non-saturated and saturated images. 

No companion with $M_{L'} = 9.8$, corresponding to the absolute magnitude of the \cc\ detected in November 2003 (if bound) is detected on the data down to a separation $r \simeq 11$--13 pixels, i.e., about 6.5~\au. Comparatively, $6\sigma$ detection limits of $M_{L'} = 9.8$ were achieved down to $r = 10$--11~pixels (i.e., about 5.5~\au) in the best set (set A) of the better quality November 2003 data. Taking into account the error bars associated to the measured magnitude (0.3~mag) does not significantly change these results (actually, the slope $\d r / \d M \simeq 4$~pixels~mag$^{-1}$ at $r \simeq 12$).

Much less homogeneous results were obtained with run B1 data. Again using HR~2435 to remove the PSF halo, a $6\sigma$ detection limits of $M_{L'} = 9.8$ is achieved down to $r=12$--17~pixels, depending on the position angle. This is coherent with the lower quality of the data compared to run A data. Using \bp\ as a comparison did not improve the results. Finally, run B2 data did not allow us to remove properly the PSF halo, as the image quality had changed signficantly between the recording of \bp\ and HR~2435 images. So, no attempt was made to derive detection limits.

\begin{figure}
  \centering
\includegraphics[width=0.95\hsize]{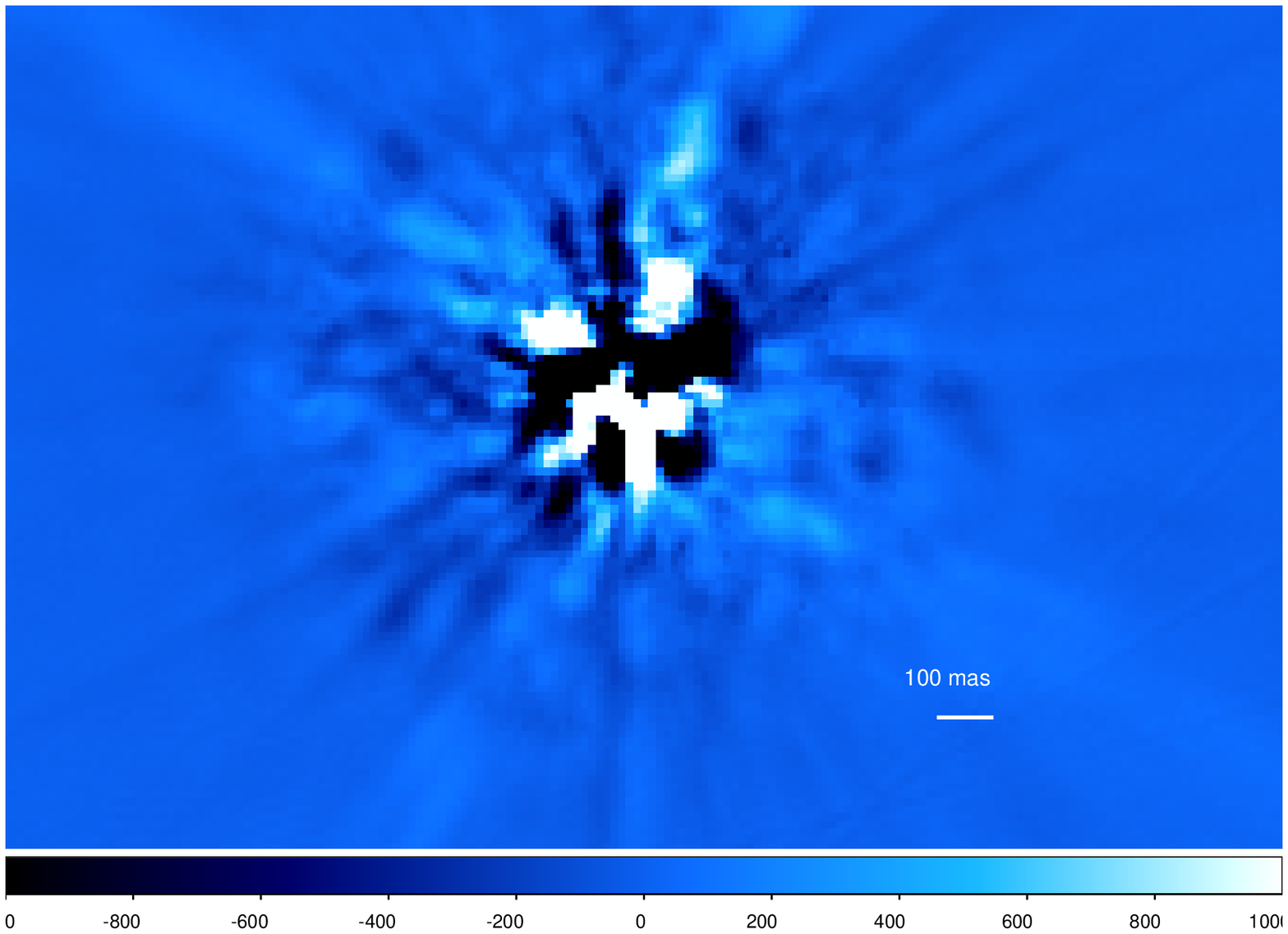}\\
\includegraphics[width=0.95\hsize]{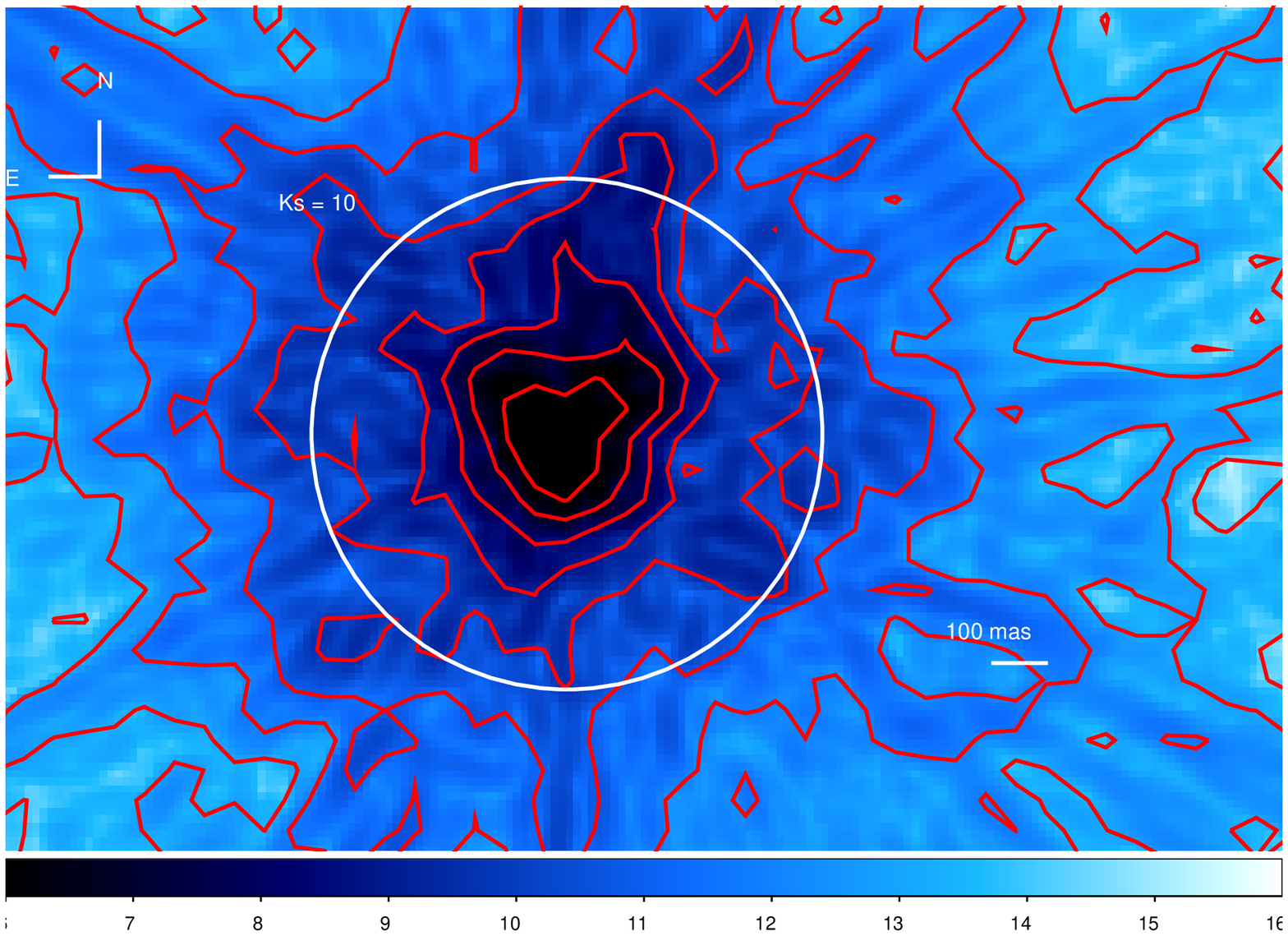}
 \caption{Top: Residual $K_s$ image after subtracting \bp\ by the comparison star HR~2435. Bottom: Map of the corresponding $6\sigma$ detection limit. The contour levels represent iso-absolute magnitudes and are separated by 1~mag. The $M_{K_s} = 10$ contour level is indicated for reference.  The \cc\ with an expected absolute magnitude $M_{K_s} \simeq 11.5$ is not detected in these data. North is up and east is to the left. The radius of the white circle corresponds to 8~\au, i.e., the projected separation of the \cc\ in November 2003.}
  \label{images_anthony}
\end{figure}

Similarly, we show in Fig.~\ref{images_anthony} the results obtained in the $K_s$ band. No companion is detected in the present data which reach a typical limit of $M_{K_s} = 9$ at a separation of 0\farcs4 (8~\au). An absolute magnitude $M_{K_s}$ of 11.3, corresponding to the one expected from the \cc\ if bound (see above) is reached further than 40--44 pixels, i.e., at about 10~\au.  Hence, the present $K_s$ data do not allow to test the presence of the \cc\ at 0\farcs4 (or less). For comparison, the computation of the $6\sigma$ detection limit in the November 2004 4QPM data show that a \cc\ with $M_{K_s} = 11.3$ would have been detected down to about 6.5--7~\au. 

\section{Constraints on the physical position of the companion candidate}
\subsection{Constraints from available deep images}
\begin{figure}
  \centering
\begin{tabular}{cc}
\includegraphics[width=0.9\hsize]{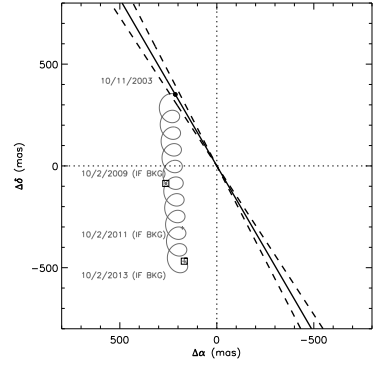}
 \end{tabular}
  \caption{Expected position of the source detected in November 2003 (black dot), assuming it was a background (BKG) object, taking the parallactic and stellar proper motions into account, for the 10 following years (thin curly line). Note that we assume, as usually done, that the \cc\ has no proper motion. The \bp\ debris disk midplane is indicated by the thick plain line. The thick dashed lines represent the uncertainty on the midplane angle.}
  \label{bkg}
\end{figure}

First, we remind that as expected, the non-detection of the \cc\ in February 2009 does not bring new constraints on the background object scenario. This can be seen in Fig.~\ref{bkg} where we have plotted the position of the \cc\ between 2003 and 2013, taking into account the stellar proper motion. Note that we assume, as usually done, that the \cc\ has no proper motion). Should the object be a physical companion, our new non-detection would place strong constraints on its possible orbits; in the following we discuss this scenario.

We will assume that the $M_{L'} = 9.8$ \cc\ orbits within the plane of the debris disk itself; the orbit is then seen edge-on as well; we furthermore assume that its orbit is prograde relative to the disk (see, e.g., \cite{olofsson01}), and circular. Assuming a circular orbit is justified by the fact that the models proposed so far to explain the \bp\ disk peculiarities (disk assymetries, FEBs) require planets on circular or low eccentric orbits (i.e., $\leq 0.1$; see a review in Lecavelier des Etangs \& Vidal-Madjar 2009), and as shown by the same authors, assuming such a low eccentricity does not change significantly the dynamical results. Besides, the relevant parameters adopted for the star itself are its mass, $1.75\pm0.05$~\Msun, and its distance $19.3\pm0.2$~pc (\cite{crifo97}). 

\begin{figure}
  \centering
\begin{tabular}{c}
\includegraphics[width=0.95\hsize]{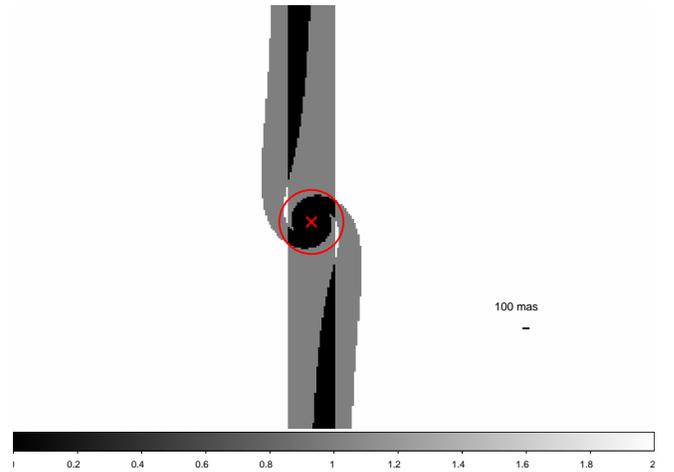}
 \end{tabular}
  \caption{Coverage map of the \bp\ system as seen from above (pole-on). Regions (as seen 
in November 2003) which correspond to detection limits enough to detect the \cc\ and 
which have been visited both in 2003 and 2009 (white area), once, ie either in 2003 
or in 2009 (grey) and which have not been visited yet (black). 
\bp\ is indicated by a star. The circle represents an orbital radius of 8 \au\ centered on the 
star. The disk rotates clockwise and 
the \cc\ was 8 \au\ on the left side of the star in November 2003 data. The 
two vertical lines correspond to separations of -6.5 and 6.5 \au. For comparison, with the 
February 2009 (resp. November 2003) data only, the whole region ``in front of'' 
and `behind' the star at separations smaller than 6.5 \au\ (resp. 5.5 \au) were 
totally un-explored. Observing at two different epochs has allowed to explore most of this 
region as well, thanks to the keplerian rotation. }
  \label{orbite_2D}
\end{figure}

In Fig.~\ref{orbite_2D}, we show the various regions of the \bp\ disk that have been explored in November 2003 and/or in 2009. With only a single epoch of observation, an important part of the star immediate surroundings is not explored. The new observations, in contrast, allow to significantly increase the explored surroundings. We see, in particular, that the full 8--10~\au\ annulus around the star has now been fully explored, i.e., a companion with a mass higher than a few Jovian masses with a separation in the 8--10~\au\ range would have been detected either in 2003 or 2009 (circular orbit assumed). In other words, we can now exclude the presence of a massive \cc\ at these separations apart from the one detected in 2003.

\begin{figure}
  \centering
\includegraphics[width=0.95\hsize]{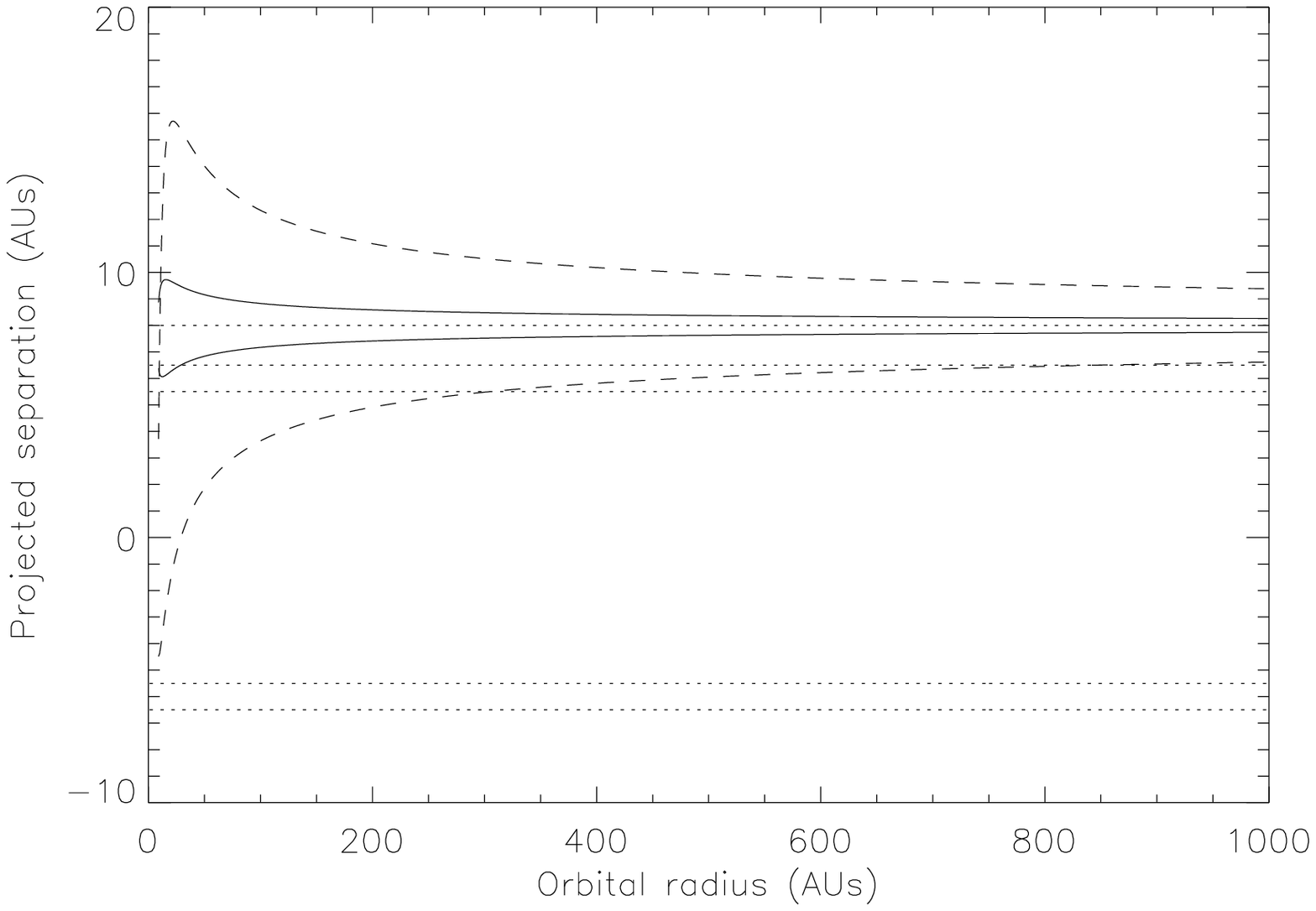}
\includegraphics[width=0.95\hsize]{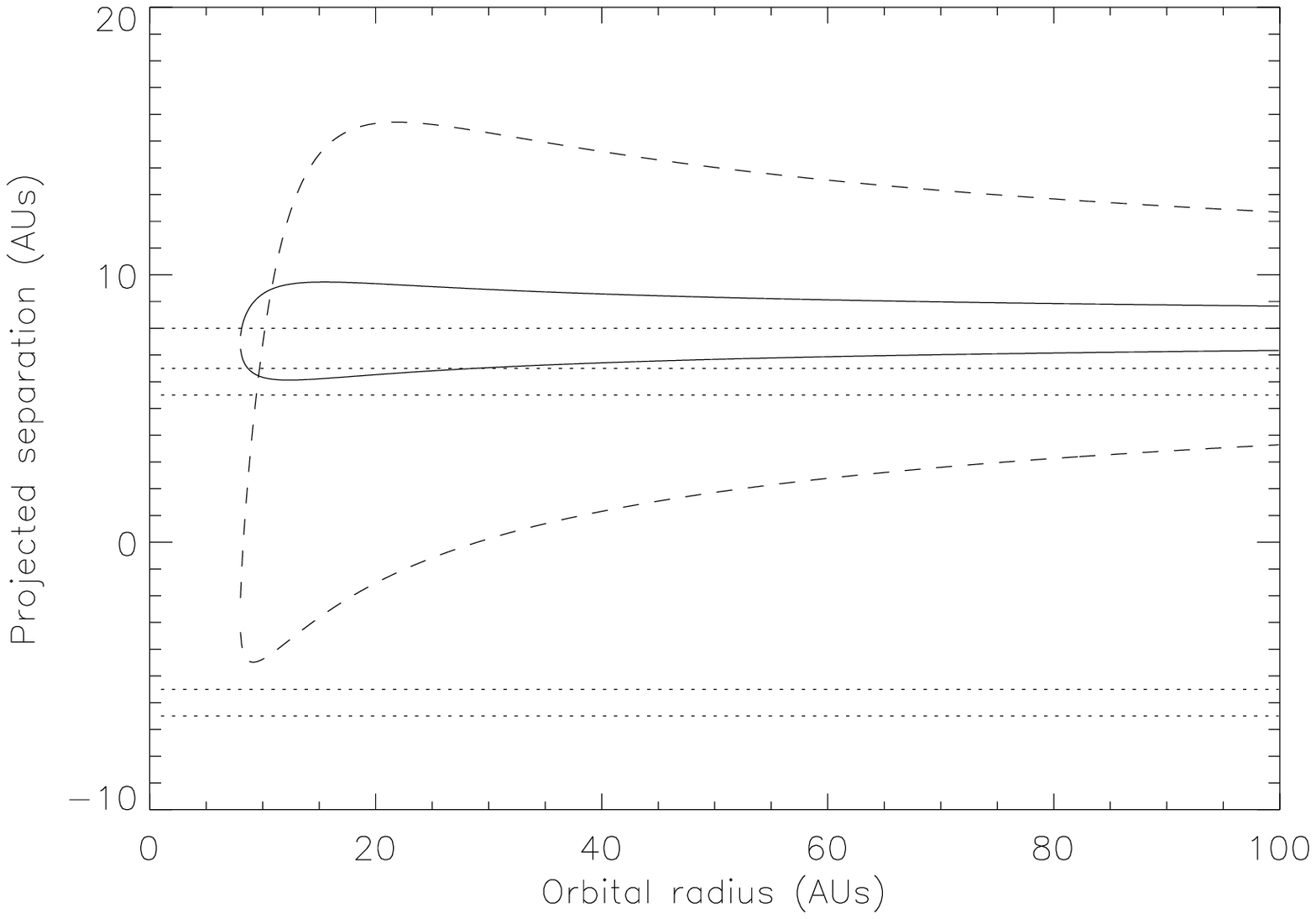}
  \caption{Projected separations of the \cc\ in 2009 (dashed curve) and 2004 (solid curve), 
as a function of its orbital radius, assuming its projected separation was 8\au\ in 2003. 
Top: orbital radii up to 1000 \au\ are considered. Bottom: zoom of the same plot, with 
orbital radii up to 100 \au. The horizontal dotted lines indicate separations  
corresponding to the computed 6 sigma detection limits of an \Lpabs = 9.8 source in 
November 2003 (5.5 \au\ ), and to an \Lpabs = 9.8 source in February 2009 (6.5 \au\ ) or 
a \Ksabs =11.3 source in November 2004 (6.5 \au\ ), and the \cc\ projected 
separation in November 2003 (8 \au). Circular orbits are assumed and the \cc\ is assumed 
to rotate prograde with respect to the circumstellar disk.}
  \label{orbite}
\end{figure}

We now constrain the physical position of the \cc\ knowing that its projected separation was 8~\au\ in November 2003 and is less than 6.5~\au\ in February 2009. In Fig.~\ref{orbite}, we plot its projected separation in February 2009, assuming its projected separation was 8~\au\ on November 2003, as a function of its orbital radius. We note that the impact of the error associated to the star mass is negligible ($\leq 3$\%). We can see that:
\begin{itemize} 
\item Initial configurations where the \cc\ orbited in 2003 before quadrature (i.e., before the maximum elongation, see Fig.~\ref{transit} for an illustration) are ruled out except for separations between 8 and 9.75~\au. Larger separations are excluded. We note that to rule out a \cc\ before quadrature and with a separation between 8 and 9.5~\au\ with the $L'$ data, we will need to wait until 2012-2013.  
\item Initial configurations where the \cc\ orbited in 2003 after quadrature lead to current positions too close to the star to be detectable in February 2009. However, we note that for orbits with radii between 8 and 10~\au, the projected separation is about 4~\au\ ( $\simeq 8$~pixels), so not far from the detection limits in case of very good atmospheric conditions. 
\end{itemize}

If we now take into account the non-detection of a \cc\ with $M_{K_s} = 11.3$ down to 6.5--7.0~\au\ in the 4QPM November 2004 data, we see (Fig.~\ref{orbite}) that (1) any \cc\ orbiting before quadrature in 2003 would have been easily detected in 2004, (2) a \cc\ orbiting after quadrature in 2003 with an orbital radius larger than 100~\au\ would also have had in 2004 a separation large enough to be easily detected (even assuming the largest possible magnitude for the \cc\ given the error bars, i.e., $M_{K_s} = 11.5$), (3) a \cc\ orbiting after quadrature in 2003 with an orbital radius in the range 8.5 to 30~\au\ would definitely not have been detected in the 4QPM 2004 data, and (4) a \cc\ orbiting after quadrature in 2003 with an orbital radius in the range 30--100~\au\ or smaller than about 8.5~\au\ would have had a separation in the range 6.5--7.1~\au\ in November 2004; given the uncertainties on both the $K_s$ photometry and the uncertainty associated to the detection limits themselves, we cannot claim that the \cc\ would have been detected in 2004.
 
We conclude then that the 2004 data definitely rule out any \cc\ which would have been before quadrature in 2003, as well as \cc\ after quadrature in 2003, with large orbital radius larger than 100~\au. The data do not allow to derive conclusions for \cc\ orbital radius in the range 8 to 30--100~\au. We remind that the conclusions derived from the 4QPM data rely on $K_s-L'$ color estimations. Knowing the actual $K_s$ magnitude of the \cc\ would therefore be very important.

\subsection{Can the companion candidate still be responsible for the 1981 photometric eclipse?}
Lecavelier des Etangs \& Vidal-Madjar (2009) proposed that in order to be responsible for the 1981 eclipse, the \cc\ observed in 2003 should have a physical separation in the range 7.6--8.7~\au\ (see Sect.~1). As underlined, they could not at that time disentangle a \cc\ which was before or after quadrature in 2003. There was thus a degeneracy in the solutions.

\begin{figure}
  \centering
\includegraphics[width=0.95\hsize]{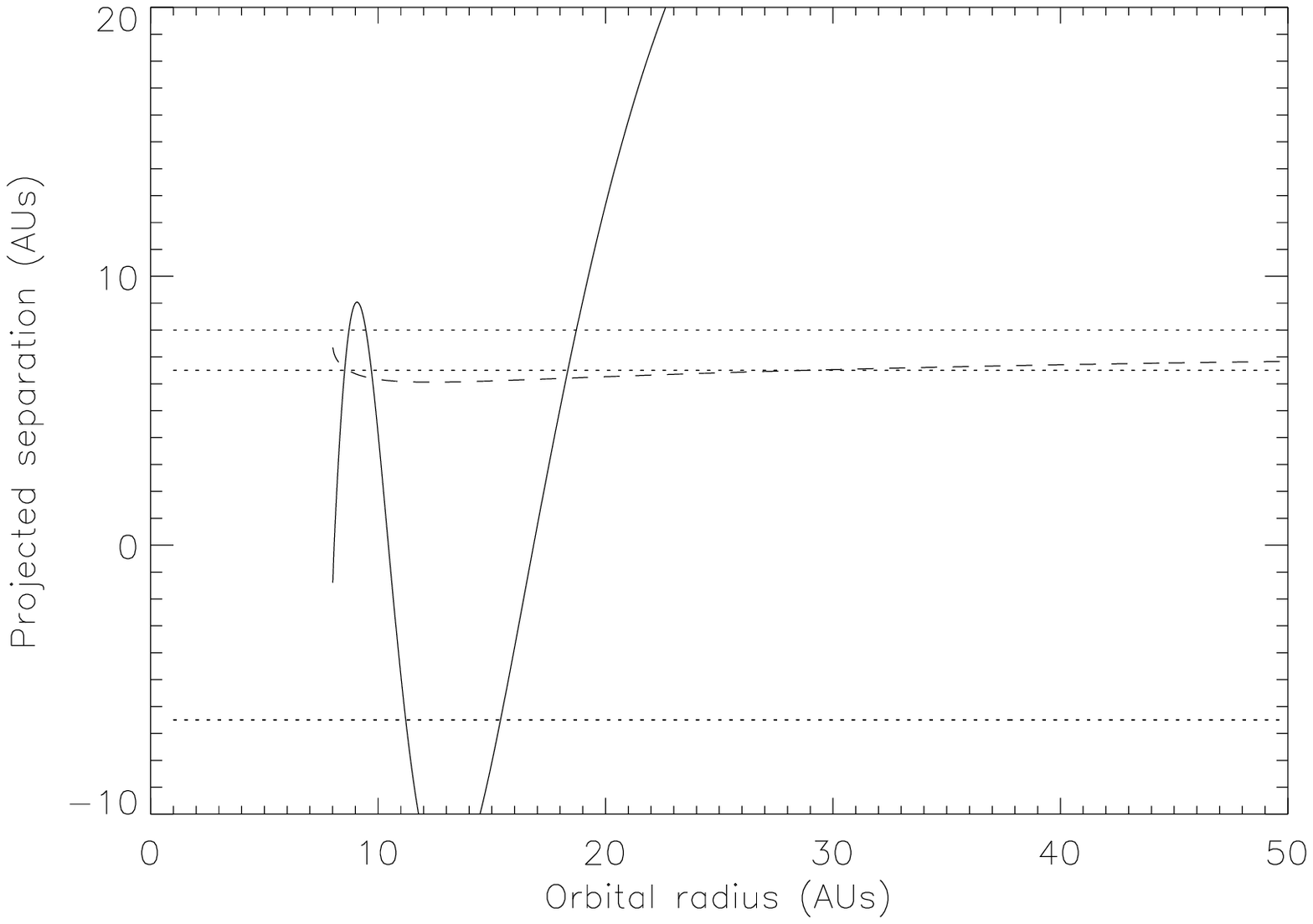}\\
\includegraphics[width=0.95\hsize]{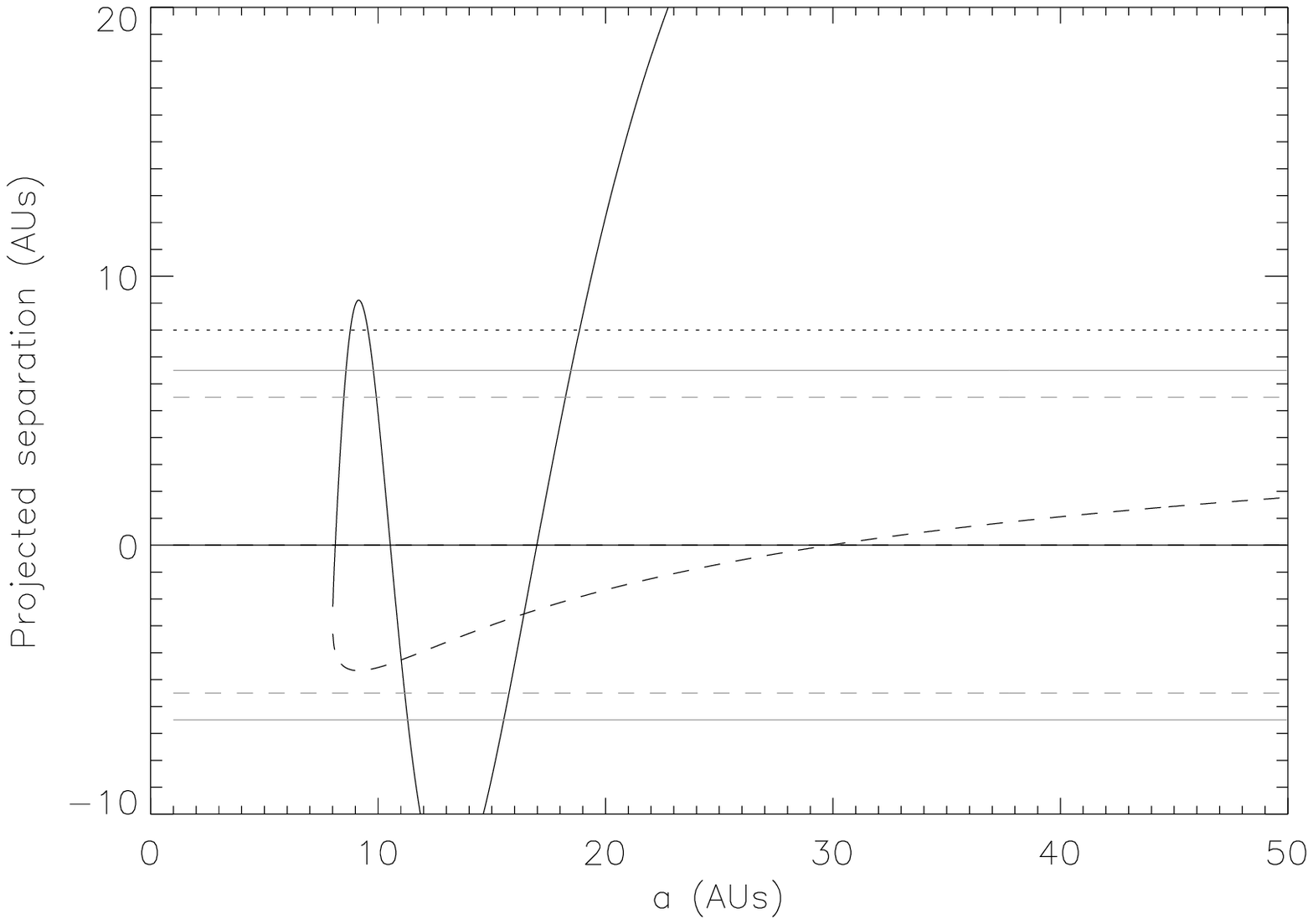}
  \caption{Top: Projected separation of the \cc\ in 2004 (dashed curve) and 1981 (solid line curve), as a function of its orbital radius, assuming its projected separation was 8 \au\ in 2003. Bottom:  Projected separation of the \cc\ in 2009 (dashed curve) and in 1981 (solid line curve), as a function of its radius, assuming its projected separation was 8 \au\ in 2003. In both cases, we took into account the fact that the \cc\ was located after quadrature in November 2003. Horizontal lines: same conventions as in Fig.~\ref{orbite}. Circular orbits are assumed and the \cc\ is assumed to rotate prograde with respect to the circumstellar disk.}
  \label{orbite_m22}
\end{figure}

We now test whether the constraints brought by the 2004 $K_s$ and 2009 $L'$ images still allow the \cc\ to be the transiting planet of 1981. The 2004 data excludes now any \cc\ which would have been before quadrature in 2003. This removes the degeneracy for the possible locations of the \cc\ in November 2003. Assuming that the \cc\ was at a projected separation of 8 \au\ in November 2003, and was after quadrature, we plot in Fig.~\ref{orbite_m22} the positions of the \cc\ back in 1981. Three possible orbital radii are compatible with a projected separation of 0 \au\ in 1981: 8.1, 10.5 and 17 \au. The radii at 10.5 \au\ corresponds to an anti-transit solution (where the \cc\ is aligned with the star and the observer, but is located exactly behind the star), so is not acceptable. The 8.1 and 17 \au\ orbital radii correspond to transit positions in 1981 and are therefore dynamically acceptable. 

\begin{figure}
  \centering
  \begin{tabular}{cc}
    \includegraphics[width=0.95\hsize]{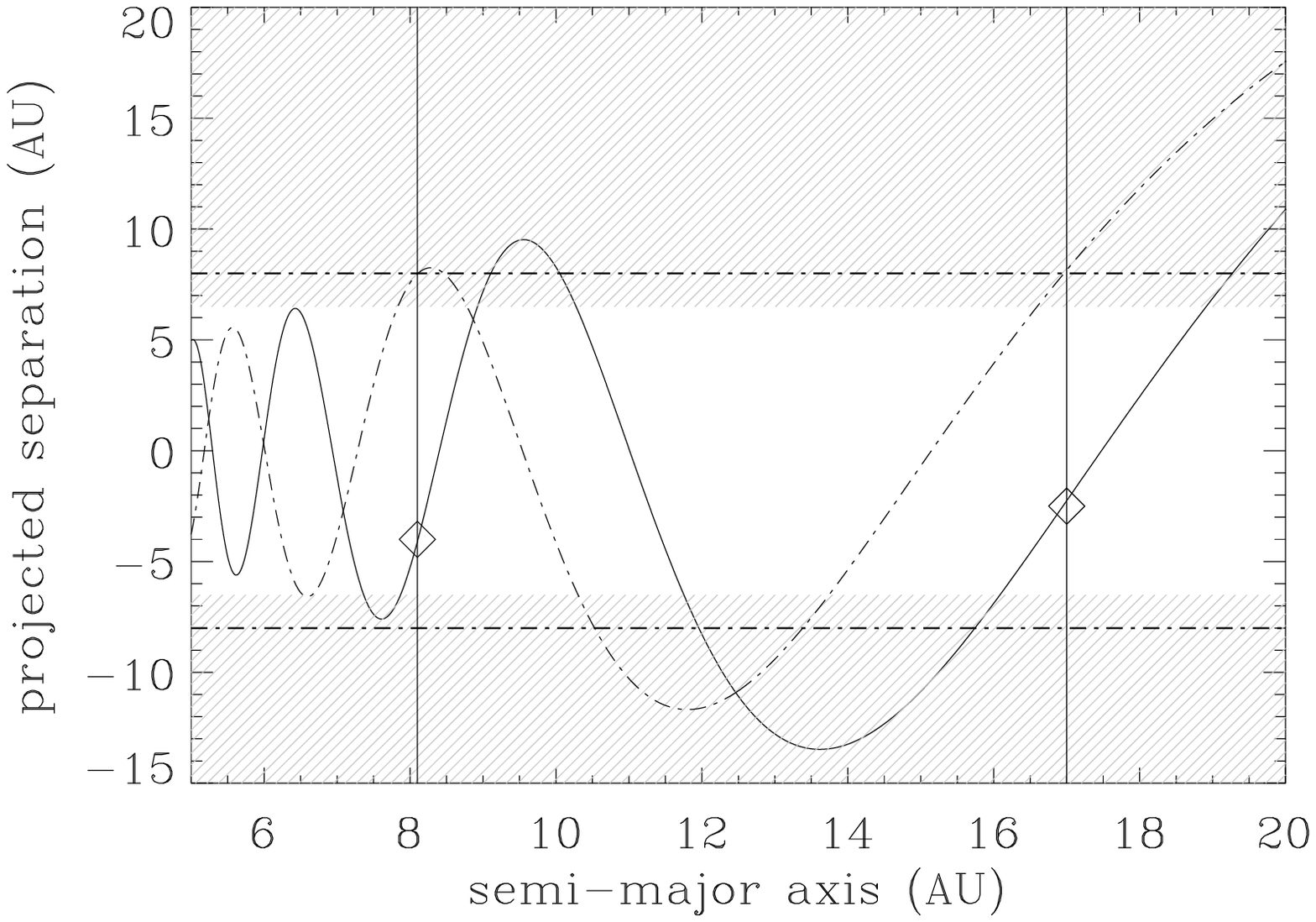}\\
    \includegraphics[width=0.75\hsize]{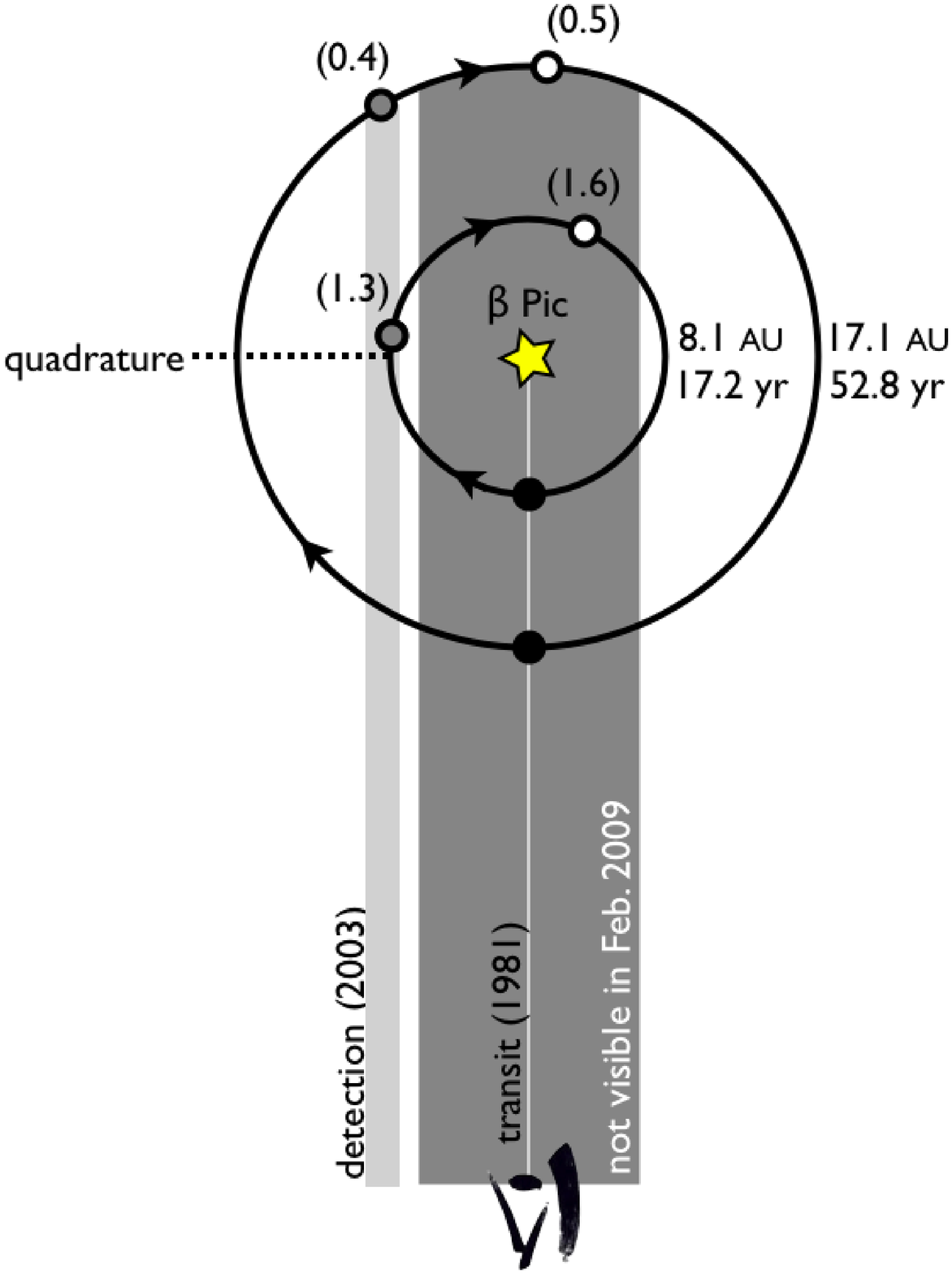}
  \end{tabular}
  \caption{Top: Projected separation of the \cc\ in 2003 (dash-dotted curve) and 2009 (solid curve) 
as a function of its orbital radius, assuming it was transiting \bp\ in 1981. 
The horizontal dash-dotted line represent the projected separation of the \cc\ in 2003.
Circular orbits are assumed, and the \cc\ is assumed to rotate prograde with respect to the circumstellar disk.
Bottom: Sketch of the possible locations in November 2003 and February 2009 (as seen from above) 
of the \cc\ assuming it transited \bp\ in 1981. The two cases a= 8.5 and a=17.1 \au\ are represented 
(see text). The values in parenthesis give the fraction of the orbit that has been covered 
in 2003 (or 2009).}
\label{transit}
\end{figure}

An alternative (but equivalent) approach, as adopted by Lecavelier des Etangs \& Vidal-Madjar (2009) is to consider the projected separation of the \cc\ in 2003 and 2009, assuming that it was transiting in 1981. In that case, we compute the projected separation assuming that at a time $t=0$, the \cc\ is transiting, where as in the previous approach, we assumed that at $t=0$, the \cc\ had a projected separation of 8~\au. The results are shown in Fig.~\ref{transit}. We see that the projected separation of the \cc\ in 2003 is compatible with the detected (projected) position at 8~\au\ for semi-major axis of 8.1, 8.5, and 17.1~\au. The orbits with semi-major axis of 8.1 and 8.5~\au\ correspond to positions before and after quadrature, respectively. The before-quadrature scenario is excluded by the 2004 4QPM images (see Fig.~4, right panel). We furthermore see that with the 8.1, and 17.1~{\sc au} solutions, the projected separation of the \cc\ in February 2009 is much smaller than 6.5~{\sc au} (unhatched region in top panel of Fig.~\ref{transit}). We conclude then that both possible semi-major axis of 8.5 and 17.1~\au\ are compatible with the \cc\ being responsible for the 1981 eclipse,  and being at a projected separation of 8~\au\ after quadrature in November 2003, and being undetected in February 2009. 

We note that a \cc\ with an orbital radius of 8.1~\au\ was at the limit of detectable zone in 2004 in the 4QPM images, but as seen earlier, it is not possible from these data to definitively draw any firm conclusions on the detectability of a \cc\ with a $\leq 30$~\au\ in 2004. We therefore regard the  8.1~\au\ solution as a possible one. A \cc\ with an orbital radius of 17.1~\au\ was much too close to the star to be detected in 2004 or 2009. Lecavelier des Etangs \& Vidal-Madjar (2009) proposed that a planet orbiting at $17.1$~\au\ could have been detected when it was near quadrature between 1993 and 1998, and thus excluded this solution. This is in fact not so clear, as the visible magnitude of the \cc\ is estimated to $V = 21$, below the detection limits of the 1997 \emph{HST}/STIS observations, that probed the surroundings of \bp\ as close as 15~\au\ for companions with $V < 17$ (Heap et al. 2000; Lagrange et al. 2009). 

\subsection{Expected separations of the \cc\ in forthcoming years}
\begin{figure}
  \centering
  \begin{tabular}{cc}
    \includegraphics[width=0.9\hsize]{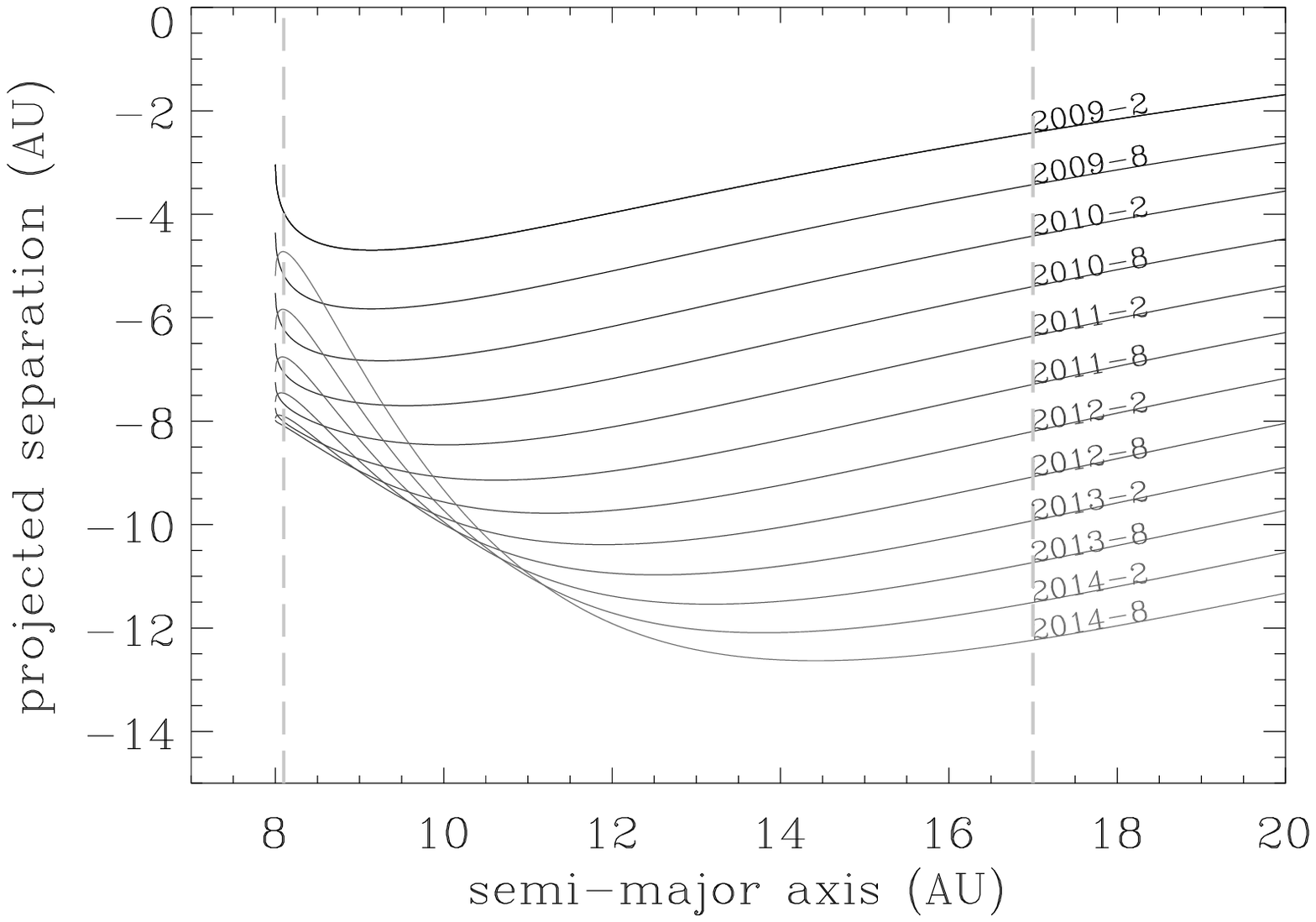}\\
    \includegraphics[width=0.9\hsize]{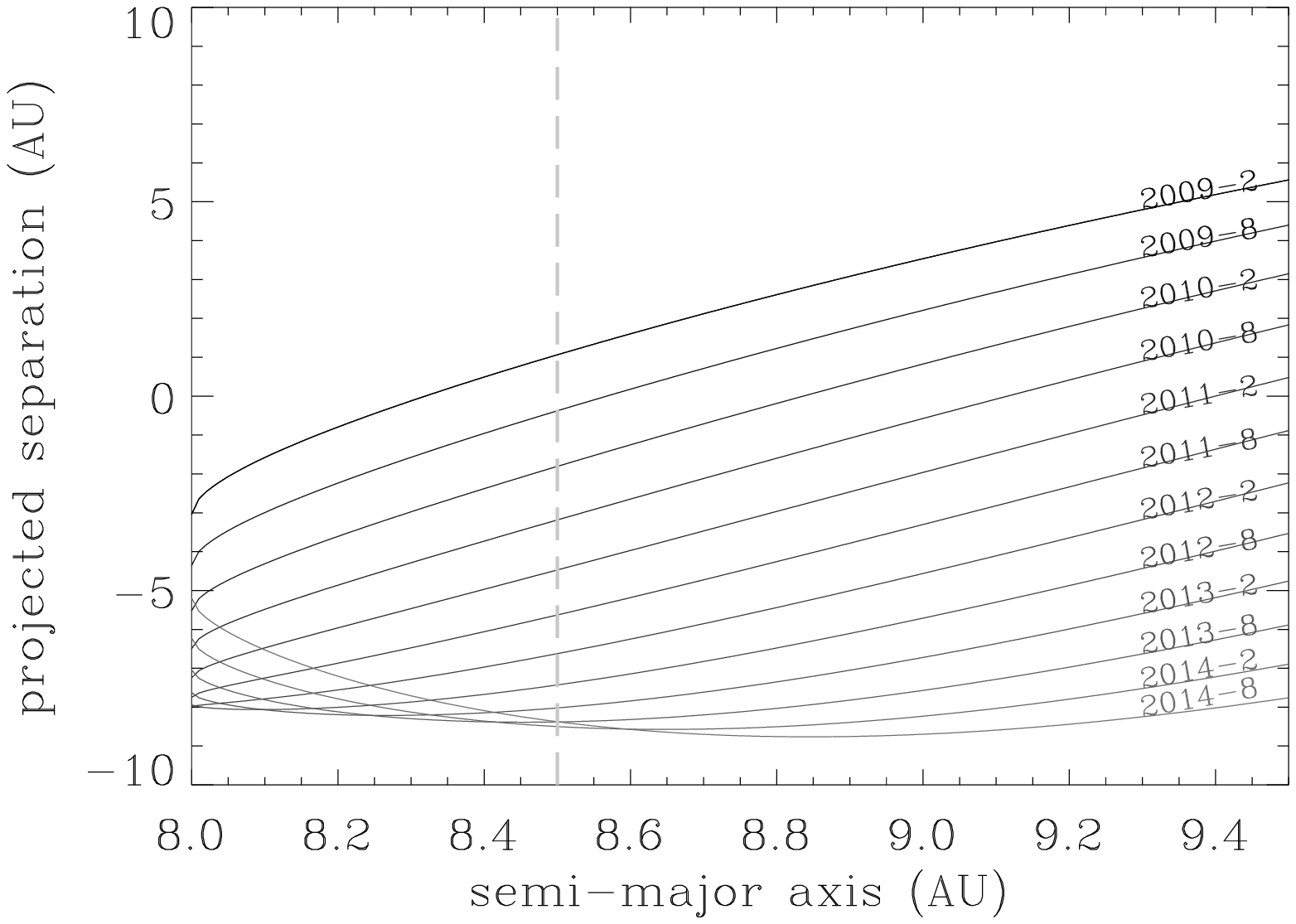}\\
  \end{tabular}
  \caption{Projected separation of the \cc\ in forthcoming years as a function of its 
orbital radius, assuming its projected separation was 8\au\ in 2003. Left: we assume, as suggested by the Ks-band images, that the \cc\ was located after quadrature in November 2003. Right: projected separation if the \cc\ was observed \emph{before} quadrature in November 2003. The vertical dashed lines at 8.5 and 17.1~{\sc au} indicate the possible (circular) separations if we further assume that the \cc\ transited in 1981. }
\label{orbite_2_new}
\end{figure}

Given the constraints brought by the $L'$ and 4QPM images, we plot in Fig.~\ref{orbite_2_new} the projected separation of the \cc\ in the forthcoming years, assuming a circular orbit. We see that if its actual orbital radius is low (between 8 and 12--15~\au), the \cc\ should be detectable again in fall 2009, under good atmospheric conditions. If we take into account only the constraints provided by $L'$-band data, we cannot exclude a \cc\ before quadrature with an orbital radius smaller than 9.5~\au; in that case (see also Fig.~\ref{orbite_2_new}), the \cc\ would be detectable again after 2012 only. 

\section{Summary and future prospects}
New $L'$-band data were obtained in February 2009 and did not reveal the presence of the \cc\ detected in 2003 down to 6.5~\au. This non-detection does not allow to rule out a background companion. The $L'$-band data allow to exclude initial positions for the \cc\ before quadrature and with radii larger than 9.75~\au\ (assuming circular orbits). They, however, do not constraint the possible orbits of a \cc\ located after quadrature in 2003. 4QPM data obtained in January 2009 did not have the sensitivity to detect the \cc\ at $K_s$ band. 

Assuming a $K_s-L'$ color of $1.2$--$1.5$~mag, a similar analysis made on the 2004 4QPM data allowed to exclude all initial configurations where the \cc\ was located before quadrature. They, moreover, restrict the possible orbital radius of the \cc\ to less than a maximum radius of $\simeq 30$--100~\au\ from the star. We remind, nevertheless, that the conclusions derived from the 4QPM data rely on $K_s-L'$ color estimations. 

Finally, we have shown that if its actual orbital radius is small (between 8 and 12--15~\au), the \cc\ could be detectable at $L'$ as soon as fall 2009 under very good atmospheric conditions in case it was seen after quadrature in 2003, and after 2012 in case it was seen before quadrature. 

\begin{acknowledgements}
We would like to thank ESO staff, in particular Lowell Tacconi for his very efficient support during phase II preparation, Paranal staff for the SM observations, and Christophe Dumas for his help. We thank Z.~Haiman for pointing out precious information on galaxies/QSO contamination. We acknowledge financial support from the French Programme National de Plan\'etologie (PNP, INSU), as well as from the French Agence Nationale pour la Recherche (ANR; project $NT05-4$\_$44463$). These results have made use of the SIMBAD database, operated at CDS, Strasbourg, France. 
\end{acknowledgements}

\end{document}